\input{aipcheck}

\documentclass[
    ,final            
  ]
  {aipproc}

\layoutstyle{6x9}

\begin{document}

\title{A deep look at the inner regions of the mini-BAL QSO PG~1126-041 with XMM-\textit{Newton}}


\author{Margherita Giustini}{
  address={Department of Astronomy \& Astrophysics, Penn State University, University Park, PA 16802, USA}
,altaddress={Dipartimento di Astronomia, Universit\`a di Bologna, via Ranzani 1, I-40127 Bologna, Italy}
}

\author{Massimo Cappi}{
  address={INAF/Istituto di Astrofisica Spaziale e Fisica cosmica, via Gobetti 101, I-40138 Bologna, Italy}
}

\author{George Chartas}{
  address={Department of Physics and Astronomy, College of Charleston, Charleston, SC 29424, USA}
  ,altaddress={Department of Astronomy \& Astrophysics, Penn State University, University Park, PA 16802, USA} 
}

\author{Michael Eracleous}{
  address={Department of Astronomy \& Astrophysics, Penn State University, University Park, PA 16802, USA}
}

\author{Giorgio G.C. Palumbo}{
  address={Dipartimento di Astronomia, Universit\`a di Bologna, via Ranzani 1, I-40127 Bologna, Italy}
}

\author{Cristian Vignali}{
  address={Dipartimento di Astronomia, Universit\`a di Bologna, via Ranzani 1, I-40127 Bologna, Italy}
}

\begin{abstract}
 A long XMM-\textit{Newton} observation of the mini-BAL QSO PG~1126-041 allowed us to
 detect a highly ionized phase of X-ray absorbing gas outflowing at $v\sim 15\,000$~km~s$^{-1}$. Physical implications are briefly discussed.
 \end{abstract}

\maketitle


\section{Introduction}
Mini-Broad Absorption Line Quasars (mini-BAL QSOs) show highly blueshifted (up to considerable fractions of the speed of light) ultraviolet absorption lines with FWHM usually ranging between 500 and 2000 km~s$^{-1}$, i.e. narrower than those of the classical BAL QSOs. While these spectral features do provide evidence for the presence of outflows from the nuclear regions of such AGNs, the physical scenario in which these phenomena fit is still not clear. X-ray observations can help contraining the physical mechanism responsible for the launch and acceleration of such outflows.

\section{The XMM-Newton long look}
XMM-\textit{Newton} observed the mini-BAL QSO PG~1126-041 ($z\sim 0.06$, UV outflow velocity $v_{\rm{uv}}\sim 5\,000$~km~s$^{-1}$) for $\sim 130$~ks on 06/21/2009. This observation provided a very high signal-to-noise ratio spectrum.

\begin{figure}[h!]
  \includegraphics[height=.25\textheight]{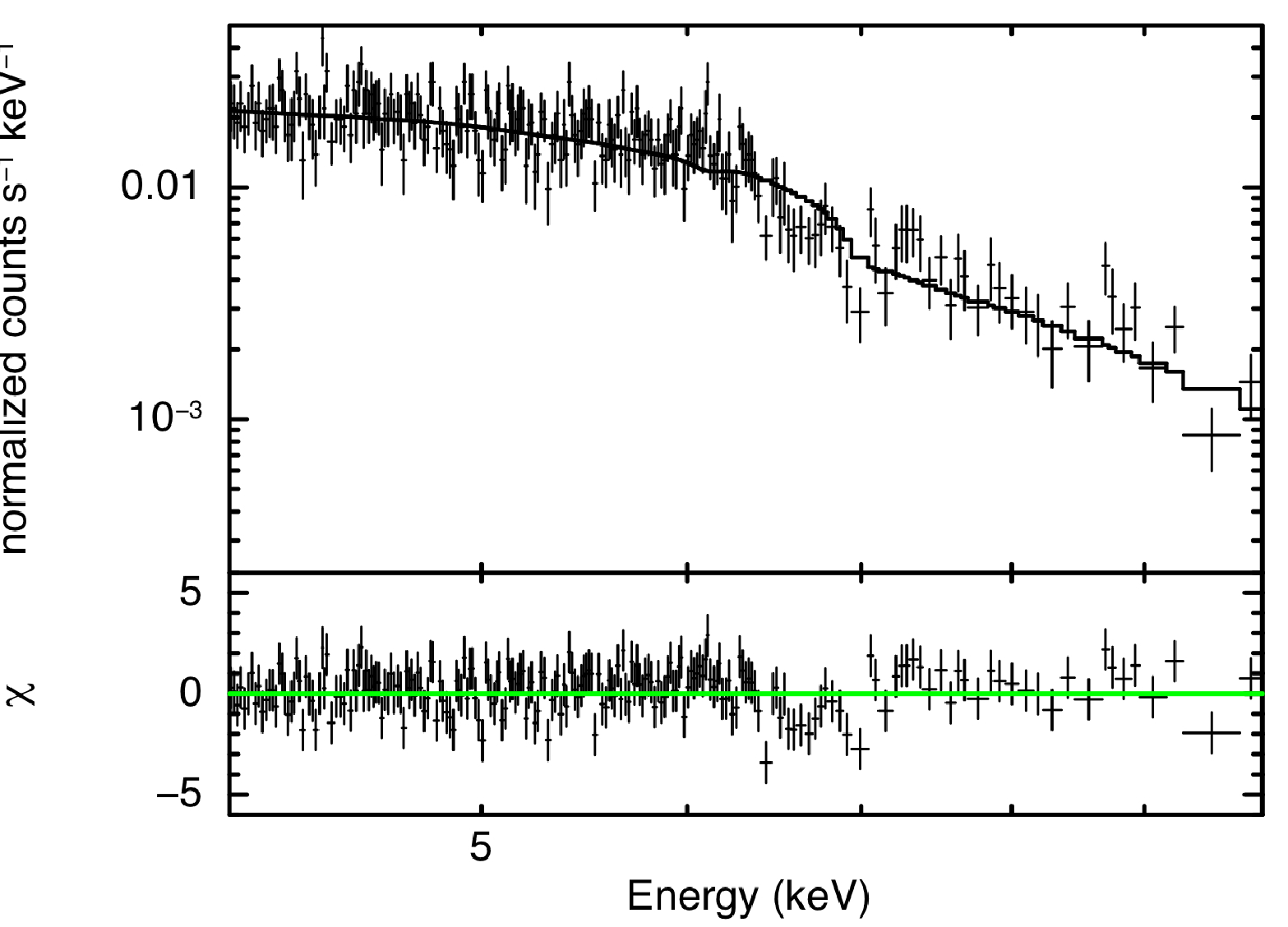}
  \includegraphics[height=.25\textheight]{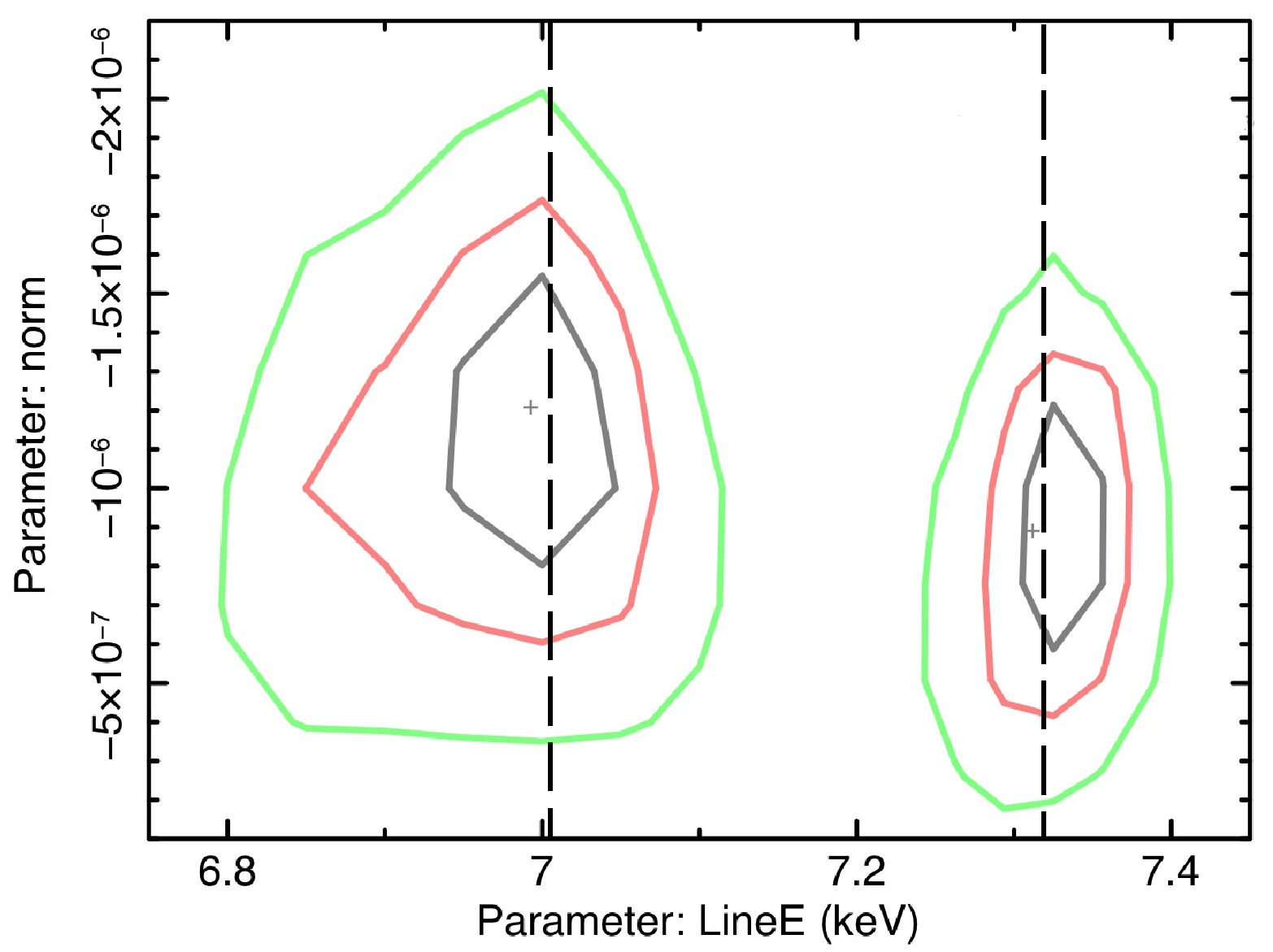}
  \caption{\label{FIG}Left: 4-10 keV EPIC-pn spectrum modeled with a powerlaw partially covered by an ionized absorber; 
  the data residuals in unit of $\sigma$ are reported in the bottom panel. Right: 68, 90 and 99\% confidence level contours for the normalization and centroid energy of the two gaussian absorption lines included in the best fit model.}
\end{figure}

We focus here on reporting the main results of the spectral analysis performed on the hard X-ray spectrum, i.e. in the 4-10~keV band. 
Fig.\ref{FIG} (left panel) shows the EPIC-pn spectrum modeled with a powerlaw with photon index $\Gamma\sim 2.5$ partially covered ($\sim 95\%$ covering factor) by a ionized absorber with a ionization parameter $\xi\sim 100$~erg~cm~s$^{-1}$ and a column density $N_H\sim 3\times 10^{23}$~cm$^{-2}$. 

The addition of two narrow Gaussian absorption lines at rest frame energies $E_1\sim 7.01$~keV and $E_2\sim 7.33$~keV significantly improves the fit ($\Delta\chi^2/dof =46/6 $). In Fig.\ref{FIG} (right panel) we report the 68, 90, and 99\% confidence level contours for the absorption lines centroid energies and normalizations. The dashed lines mark the \textit{expected} centroid energy of FeXXV~K$\alpha$ and FeXXVI~K$\alpha$ transitions, both blueshifted by $v=0.05c$. The correspondence with the \textit{detected} centroid energies is striking.

The two iron absorption lines can be reproduced by a self-consistent photoionization model using \textsc{XSTAR} \cite{Xstar}. This requires the addition of another ionized absorber, with a ionization parameter $\xi\sim 1900$~erg~cm~s$^{-1}$ and an outflow velocity $v_{\rm{x}}\sim 0.05c$. The ionization parameter is tightly constrained given the presence of both FeXXV~K$\alpha$ and FeXXVI~K$\alpha$ transitions with similar equivalent widths (EW$\sim 95$~eV). The column density $N_H \approx 10^{23-24}$~cm$^{-2}$ is not well constrained.



This phase of X-ray absorbing gas is outflowing at $v_{\rm{x}}\sim3\times v_{\rm{uv}}$.


\section{Conclusions}
The two detected phases of X-ray absorbing gas are qualitatively consistent with radiation-driven accretion disk wind models predictions \cite{proga}, where one expects a high column density of X-ray absorbing gas shielding the UV-accelerated wind, and a X-ray component of the wind, much faster than the UV one. An interesting future developement would be the simulation of realistic X-ray spectra expected from accretion disk wind models to check whether they can self-consistently and quantitatively reproduce the observed spectra.


\begin{theacknowledgments}
We acknowledge financial support from NASA grant NNX08AB71G and ASI contract I/088/06/0. ME acknowledges support from NSF grant AST-0807993.
\end{theacknowledgments}



\bibliographystyle{aipprocl} 



\end{document}